\documentclass[preprint,12pt]{elsarticle}

\usepackage{amssymb}

\journal{Physics Letters B}

\begin{document}

\begin{frontmatter}




\title{Thermodynamics, geometrothermodynamics  and  critical
    behavior of (2+1)-dimensional black holes}

\renewcommand{\thefootnote}{\fnsymbol{footnote}}

\author{ YiWen  Han,  Gang Chen}
\address{College of Computer Science, Chongqing
Technology and Business University,  Chongqing 400067,  China }

\begin{abstract}
In this paper, we study the properties of the (2+1)-dimensional
black holes from the viewpoint of geometrothermodynamics. We show
that the Legendre invariant metric of the (2+1)-dimensional black
holes can produce correctly the behavior of the thermodynamic
interaction and phase transition structure of the corresponding
black hole configurations. We find that they are both curved and the
curvature scalar gives the information about the phase transition
point.
\end{abstract}


\end{frontmatter}


\section{Introduction}

The black hole thermodynamics has been one of the focuses in
theoretical physics during the past thirty years [1-9]. The results
showed that a black hole is a thermodynamics system, it has Hawking
temperature proportional to its surface gravity on the horizon, and
they satisfy the four laws of black hole thermodynamics. However, in
geometry framework, black hole thermodynamics has been investigated
from the critical points of moduli space by using the Weinhold
metric and Ruppeiner metric [10]. As is well known, an interesting
inner product on the equilibrium thermodynamic space of state in the
energy representation was proposed by Weinhold as the Hessian matrix
of the internal energy $U$ with respect to the extensive
thermodynamic variables $N^a$, namely
g$_{ij}^W=\partial_i\partial_jM(U,N^a)$ [11]. However, there was no
physical interpretation associated with this metric structure. As a
modification, Ruppeiner introduced Riemannian metric into
thermodynamic system once more and defended it as the second
derivative of entropy $S$ (here, entropy is a function of internal
energy $U$ and its extensive variables $N^a$)
g$_{ij}^R=-\partial_i\partial_jS(U,N^a)$ [12]. In the next, it was
applied to all kinds of thermodynamics modes. For example, Cai and
Cho [13] gave a brief review on the geometrical method on the
thermodynamics, and applied this approach to the BTZ black hole.
Aman et. al. [14], showed curvature scalars and phase transitions of
the BTZ and the Reissner-Nordstrom. In addition,  Ruppeiner has
given a systematic discussion on how to make the correct choice of a
metric, and has also demonstrated several limiting results matching
extreme Kerr-Newman black hole thermodynamics to the 2- dimensional
Fermi gas. This shows that the connection to a 2D model is
consistent with the membrane paradigm of black holes [15-16]. Using
the Ruppeiner¡¯s thermodynamics geometry theory, one have shown that
Ruppeiner geometry can be carried out in various thermodynamic
systems [17-26]. Such as the ideal gas, the van der Waals gas and so
on. It was shown that the scalar curvature is zero and the Ruppeiner
metric is flat for the van der Waals gas. The curvature is nonzero
and diverges only after the phase transition takes place. The key of
the above problems is the thermodynamic potential, which is
generally believed to be the internal energy rather than the mass.
Above researches have shown that Weinhold¡¯s and Ruppeiner¡¯s
thermodynamic metrics are not invariant under the Legendre
transformations.

Recently, Quevedo et al. [27] present a new formalism of
geometrothermodynamics (GTD) as a geometric approach that
incorporates Legendre invariance in a natural way, and allows us to
derive Legendre invariant metrics in the space of equilibrium
states. Considering the Legendre invariant, they present a unified
geometry where the metric structure can give a well description of
various types of black hole thermodynamics [28-31]. The aim of the
application of different thermodynamic geometries is to describe
phase transitions in terms of curvature singularities. For a
thermodynamic system, it is quite interesting to investigate the
corresponding relationship between the curvature of Weinhold metric,
Ruppeiner metric, the Legendre invariant metric and the phase
transitions. In fact, above viewpoint has been applied to various
black holes [19, 26]. Of course, it is still widely believed that
the thermodynamic geometry of a black hole is still a most
fascinating and unresolved subject today. The main purpose of the
present work is to show that the Legendre invariant metric can be
used to reproduce correctly the thermodynamics of the
(2+1)-dimensional black holes. This has been analyzed previously by
using a different approach where Legendre invariance is not taken
into account [32].

The organization of the Letter is outlined as follows. In Sec. 2, we
present a (2+1)-dimensional black hole with a coulomb-like field. In
Sec.3, show geometrothermodynamics of the (2+1)-dimensional black
hole with a coulomb-like field. Sec. 4 ends up with some discussions
and conclusions. Throughout the Letter, the units $c=k_B=\hbar=1$
are used.

\section{The (2+1)-dimensional black hole with a coulomb-like field }

 The action describing the (2+1)-dimensional Einstein theory coupled with nonlinear electrodynamics is given by [33]
$$
S=\int\sqrt{\mbox{g}}(\frac{1}{16\pi})(R-2\Lambda)+L(F))d^3x,\eqno{(1)}
$$
with arbitrary, at this stage, the electromagnetic Lagrangian
$L(F)$. We are using units in which $c=G=1$. Since there is a $T$
ambiguity in the definition of the gravitational constant £¨there is
not Newtonian gravitational limit in 2+1 dimensions£© one can
maintain the factor $\frac{1}{16\pi}$ in the action to keep the
parallelism with£¨3+1£©-gravity. The variation with respect to the
metric gives the Einstein equations
$$
G_{ab}+\Lambda\mbox{g}_{ab}=8\pi T_{ab}, \eqno{(2)}
$$
$$
T_{ab}=\mbox{g}_{ab}L(F)-F_{ac}F_b^cL_{,F}, \eqno{(3)}
$$
$$
\nabla_a(F^{ab}L_{,F})=0, \eqno{(4)}
$$
where stands $L_{,F}$ for the derivative with respect to
$F=(F_{ab}F^{ab})/4$ . The nonlinear field is chosen such that the
energy momentum tensor (3) has a vanishing trace. The trace of the
tensor gives
$$
T=T_{ab}\mbox{g}^{ab}=3L(F)-4FL_{,F}. \eqno{(5)}
$$
In order to have a vanishing trace, the electromagnetic Lagrangian
is obtained as
$$
L=C|F|^{3/4}, \eqno{(6)}
$$
where $C$ is an integration constant. One can rewrite this
Lagrangian as
$$
L=C\Big|\frac{1}{2}(B^2-E^2)\Big|=1, \eqno{(7)}
$$
when referred to orthonormal local Lorentzian basis. With reference
to the paper [34], the complete solution to the above action is
given by the metric
$$
ds^2=-f(r)dt^2+f(r)^{-1}dr^2+r^2d\theta^2, \eqno{(8)}
$$
where the metric function is $f(r)$ given by
$$
f(r)=-M+\frac{r^2}{l^2}+\frac{4Q^2}{3r}. \eqno{(9)}
$$
Here $M$ is the mass, $l^2=\Lambda^{-1}$ the case
$\Lambda>0(\Lambda<0)$, corresponds to an asymptotically de-Sitter
(anti de-Sitter) space-time, $Q$ is the electric charge. From Eq.
(9), the event horizon is located at $f(r_h)=0$ and the radius $r_h$
satisfies
$$
M=\frac{r_h^2}{l^2}+\frac{4Q^2}{3r_h}. \eqno{(10)}
$$
For the extremal black hole, there exist two event horizons, the
inner event horizon and the outer event horizon. Here, we have
denoted $r_h$ as the radius of outer event horizon. From the energy
conservation law of the black hole
$$
dM=TdS+\phi dQ, \eqno{(11)}
$$
Using the relation between entropy and the radius of the event
horizon, we can obtain
$$
S=4\pi r_h. \eqno{(12)}
$$
The thermodynamic temperature and electric potential can be
expressed
$$
T=(\frac{\partial M}{\partial S})_Q=\frac{S}{8l^2\pi^2}-\frac{16\pi
Q^2}{3S^2}=\frac{1}{2\pi}(\frac{r_h}{l^2}-\frac{2Q^2}{3r_h^2}).
\eqno{(13)}
$$
and
$$
\phi=(\frac{\partial M}{\partial Q})_S=\frac{32\pi Q}{3S}.
\eqno{(14)}
$$

\section{Geometrothermodynamics of the (2+1)-dimensional black hole with a coulomb-like field }

Now, we turn to the recent geometric formulation of extended
thermodynamic behavior of the (2+1)-dimensional black hole with a
coulomb-like field.

The formulation of GTD of black hole is based on the theory of
contact geometry as a framework for thermodynamics [27]. Consider
the (2n+1)-dimensional thermodynamic phase space $\mathfrak{J}$ with
the coordinates $Z^A=\{\Phi,E^a,I^a\}$ where $A=0,...,2n$ and
$a=1,...,n$. In ordinary thermodynamics, $\Phi$ corresponds to the
thermodynamic potential, and $E^a$,$I^a$ are the extensive and
intensive variables, respectively. The fundamental differential form
$\Theta$ can then be written in a canonical manner as
$\Theta=d\Phi-\delta_{ab}I^adE^b$ , where $\delta_{ab}$ is the
Euclidean metric. Considering a non-degenerate metric $G=G(Z^A)$,
and the Gibbs1-form, with $\delta_{ab}=diag\{1,...,1\}$, we obtain a
set $(\mathfrak{J},\Theta,G)$ which defines a contact Riemannian
manifold if the condition $\Theta\wedge(d\Theta)^n\neq0$is
satisfied. This arbitrariness is restricted by the condition that
$G$ must be invariant with respect to Legendre transformations. This
is a necessary condition for our description of thermodynamic
systems to be independent of the thermodynamic potential. This
implies that $T$ must be a curved manifold [27] because the special
case of a metric with vanishing curvature turns out to be
non-Legendre invariant. The Gibbs 1-form $\Theta$ is also invariant
with respect to Legendre transformations. Legendre invariance
guarantees that the geometric properties of $G$ do not depend on the
thermodynamic potential.

The thermodynamic phase space $\mathfrak{J}$ with a coulomb-like
field can be defined as a 5-dimensional space with coordinates
$Z^A=\{M,S,T,Q\}$, $A=0,...,4$. The Eq. (10) represents the
fundamental relationship  $M(S,Q)$ from which all the thermodynamic
information can be obtained. Therefore, we would like to consider a
5-dimensional phase space $\mathfrak{J}$ with coordinates
$(M,S,T,Q,\Phi)$, a contact1-form
$$
\Theta=dM-TdS-\phi dQ, \eqno{(15)}
$$
and an invariant metric
$$
G=(dM-TdS-\phi dQ)^2+(TS+\phi Q)(-dTdS+d\phi dQ). \eqno{(16)}
$$
The triplet  $(\mathfrak{J},\Theta,G)$ defines a contact Riemannian
manifold that plays an auxiliary role in GTD. We should properly
handle the invariance with respect to Legendre transformations. In
fact, for the charged black hole, a Legendre transformation involves
in general all the thermodynamic variables $M, S, Q, T$ and $\phi$.
So they must be independent from each other as they are in the phase
space. We introduce also the geometric structure of the space of
equilibrium states $\varepsilon$ in the following manner:
$\varepsilon$ is a 2-dimensional submanifold of $\mathfrak{J}$ that
is defined by the smooth embedding map $\varphi:
\varepsilon\mapsto\mathfrak{J}$, which satisfies the condition that
the ¡°projection¡± of the contact form $\Theta$ on $\varepsilon$
vanishes, namely $\varphi^*(\Theta)=0$, where $\varphi^*$ is the
pullback of $\varphi$. $G$ induces a Legendre invariant metric g on
$\varepsilon$ by means of $\varepsilon$. In principle, any
2-dimensional subset of the set of coordinates of $\mathfrak{J}$ can
be used in coordinative $\varepsilon$. For the sake of simplicity,
we will use the set of extensive variables $s$ and $Q$ which in
ordinary thermodynamics corresponds to the energy representation.
Then, the embedding map for this specific choice is
$$
\varphi:\{S,Q\}\mapsto \{M(S,Q),S,Q,\frac{\partial M}{\partial
S},\frac{\partial M}{\partial Q}\}. \eqno{(17)}
$$
The condition  $\varphi^*(\Theta)=0$ is equivalent to Eq. (11) (the
first law of thermodynamics), Eq. (13), Eq. (14) (the conditions of
thermodynamic equilibrium). Then the induced metric is obtained
$$
\mbox{g}=(S\frac{\partial M}{\partial S}+Q\frac{\partial M}{\partial
Q})(-\frac{\partial^2 M}{\partial S^2}dS^2+\frac{\partial^2
M}{\partial Q^2}dQ^2). \eqno{(18)}
$$
This metric determines all the geometric properties of the
equilibrium space $\varepsilon$. We see that in order to obtain the
explicit form of the metric it is necessary to specify the
thermodynamic potential $M$ as a function of $S$ and $Q$. In
ordinary thermodynamics this function is usually referred to as the
fundamental equation from which all the equations of state can be
derived.

Substituting Eq. (12) into Eq. (10), the mass can be obtained as the
function of the entropy $S$ and the charge $Q$ in the form
$$
M(S,Q)=\frac{S^2}{16\pi^2l^2}+\frac{16\pi Q^2}{3S}. \eqno{(19)}
$$
It has been established that the physical parameters of the
(2+1)-dimensional black hole with nonlinear electrodynamics satisfy
the first law of black hole thermodynamics.

Substituting Eq. (19) into Eq. (18), we can obtain the Legendre
metric components of the (2+1)-dimensional black hole with a
coulomb-like field as
$$
\mbox{g}_{SS}=-\frac{512\pi^3Q^4}{9S^4}-\frac{2 Q^2}{\pi
l^2}-\frac{S^2}{64\pi^4 l^4}, \eqno{(20)}
$$
$$
\mbox{g}_{QQ}=\frac{512\pi^2Q^2}{9S^2}+\frac{4S}{3\pi l^2}.
\eqno{(21)}
$$
 After some calculations, we obtain the Legendre invariant scalar curvature
$$
\mathfrak{R}_{L}=\frac{864\pi^4S^5l^4(425984\pi^6Q^4l^4+1152\pi^3Q^2S^3l^2-81S^6)}{(3S^3+128\pi^3Q^2l^2)^3(3S^3+256\pi^3Q^2l^2)^2}.
\eqno{(22)}
$$
The curved nature of the Legendre metric suggests that the
thermodynamics of the present black hole has statistical mechanics
analogue.

Now, for a given charge, the heat capacity has the expression
$$
C_{Q}=T(\frac{\partial S}{\partial
T})_Q=\frac{S(3S^2-128\pi^3Q^2l^2)}{3S^2+256\pi^3Q^2l^2}.
\eqno{(22)}
$$
Obviously, the heat capacities have the zero-points at
$3S^2=128\pi^3Q^2l^2$. Moreover, $C_Q$ changes sign and the scalar
curvature diverge at $3S^2=-256\pi^3Q^2l^2$. Therefore, there will
be a phase transition at  $3S^2=-256\pi^3Q^2l^2$.

\section{Conclusion and Discussion}

In this work we reproduced the thermodynamics properties such as
temperature and entropy of the (2+1)-dimensional black holes. We
also studied the Legendre invariant metric of the (2+1)-dimensional
black holes. The results show that GTD delivers a particular
thermodynamic metric for the (2+1)-dimensional black holes. Then we
could corroborate that the thermodynamic curvature is nonzero and
its singularities reproduce the phase transition structure which
follows from the divergencies of the heat capacity.

In addition, the thermodynamic metric proposed in this work has been
applied to the case of black hole configurations in three
dimensions. It has been shown that this thermodynamic metric
correctly describes the thermodynamic behavior of the corresponding
black hole configurations. One additional advantage of this
thermodynamic metric is its invariance with respect to total
Legendre transformations. This means that the results are
independent on the thermodynamic potential used to generate the
thermodynamic metric. In all the remaining cases, the singularities
of the thermodynamic curvature correspond to points where the heat
capacity diverges and phase transitions take place. We interpret
this result as an additional indication that the thermodynamic
curvature, as defined in GTD, can be used as measure of
thermodynamic interaction. In fact, it has been shown that in the
case of more realistic thermodynamic systems [30], the ideal gas is
also characterized by a vanishing thermodynamic curvature, whereas
the van der Waals gas generates a nonvanishing curvature whose
singularities reproduces the corresponding phase transition
structure.

Furthermore, we expect that this unified geometry description may
give more information about a thermodynamic system. We conclude that
GTD is, in general, duality invariant. Therefore, our results
support Quevedo's viewpoint.
\\\\
{\bf Acknowledgements}

We thank Chang-gun Gao for fruitful discussions. This work was
supported by the Scientific and Technological foundation of
Chongqing Municipal Education Commission (Grant No. KJ100706).
E-mail address: hanyw1965@163.com

\end{document}